\documentclass[10pt,showpacs,preprintnumbers,amsmath,amssymb]{revtex4}
\usepackage{units}
\usepackage{bbold,euler,newcent}
\usepackage{graphicx}
\usepackage{dcolumn}
\usepackage{bm,bbm,mathtools,esvect}
\usepackage{slashed}

\begin{document}

\title{\LARGE\sc A novel covariant approach to gravito-electromagnetism\\\vspace{5mm}}
\author{\tt\large SERGIO GIARDINO} \vspace{1cm}
\email{sergio.giardino@ufrgs.br}
\affiliation{\vspace{3mm} Departamento de Matem\'atica Pura e Aplicada, Universidade Federal do Rio Grande do Sul (UFRGS)\\
Avenida Bento Gon\c calves 9500, Caixa Postal 15080, 91501-970  Porto Alegre, RS, Brazil}

\begin{abstract}
\noindent 
This paper describes general relativity at the gravito-electromagnetic precision level as a constrained field theory. 
In this novel formulation, the gravity field comprises two auxiliary fields, a static matter field and a moving matter field.
Equations of motion, continuity equation, energy conservation, field tensor, energy-momentum 
tensor, constraints and Lagrangian formulation are presented as a simple and unified formulation
that can be useful for future research.

\end{abstract}

\maketitle
\tableofcontents
\section{\;\sc gravito-electromagnetic analogy}
Interest in the analogies between gravitation and electromagnetism, also called gravito-electromagnetism (GEM), has increased in recent years.  
These relationships were observed and reported during the second half of the nineteenth century, and we recommend \cite{Ruggiero:2002hz,Iorio:2010qs,Vieira:2016csi} for historical review and references. Various
approaches to GEM have recently been proposed, and we quote a non-exhaustive list of papers concerning, by way of example,  
gravitomagnetic effects \cite{Tartaglia:2003wx, Schaefer:2004qh,Ummarino:2017bvz,Behera:2017voq},
the relation of GEM to special relativity \cite{Malekolkalami:2006my,Vieira:2016csi},
tidal tensors \cite{Costa:2012cw,Costa:2012cy},
weak-field approximation \cite{Bakopoulos:2014exa,Bakopoulos:2016rkl,Mashhoon:2014jna,Farrugia:2020fcu},
the Lorentz violation \cite{Santos:2018acs,Santos:2018siy},
teleparallel gravity \cite{Spaniol:2008rt,Spaniol:2014lba,Ming:2017tna},
the Mashhoon-Theiss effect \cite{Bini:2016xqg},
quantum gravity \cite{Ramos:2010zza,Santos:2016vfa},
gravitational waves \cite{Sparano:2010bs,Cabral:2016klm},
the relation of GEM to electro-dynamics in curved spacetime  \cite{Cabral:2016yxh,Cabral:2016hpq},
gravitational field of astrophysical objects \cite{Ruggiero:2015ima,Ruggiero:2015pva},
the Sagnac effect \cite{Ruggiero:2014aya,Ruggiero:2015gha},
torsion gravity \cite{Bini:2015xqa},
the Schr\"odinger-Newton equation \cite{Manfredi:2014hna},
non-commutative geometry \cite{Malekolkalami:2013fqa},
spin-gravity coupling \cite{Mashhoon:2013jaa},
gravity and thermodynamics \cite{Acquaviva:2018mqb},
the Casimir effect \cite{Quach:2015qwa}, 
gauge transformations \cite{Ramos:2018rdw} and,
quantum field gravity \cite{Behera:2003md,Behera:2018prr}. 
It is commonly known that GEM is a source of  new ideas and a guide for research into new physics. Several experimental attempts
to ascertain this are reviewed in \cite{Vieira:2016csi}.

Following  chapter 3 of  \cite{Ohanian:1995uu}, we introduce gravito-electromagnetism by way of the weak-field approximation of general relativity,
where the metric tensor $g_{\mu\nu}$ reads
\begin{equation}\label{g001}
g^{\mu\nu}=\eta^{\mu\nu}+\kappa h^{\mu\nu},\qquad\qquad\textrm{with}\qquad\qquad \kappa=\frac{\sqrt{16\pi G}}{c^2}
\end{equation}
given as a constant in units of cgs and $\,h^{\mu\nu}\,$ as a perturbation of the Minkowski metric tensor $\,\eta^{\mu\nu},\,$ from the plane 
space,  
where $\,\eta^{00}=1,\,$ $\eta^{ii}=-1\,$ and $i,\,j=\{1,\,2,\,3\}$. Of course, the approximation imposes $\,|\kappa h_{\mu\nu}|\ll 1\,$. In this approach, the velocity $\,\bm v\,$ of a particle in a 
gravitational field satisfies an acceleration law that is analogous to the electro-dynamical Lorentz force law:
\begin{equation}\label{g002}
\frac{d\bm v}{dt}=\bm g+\bm{v\times b},
\end{equation}
where $\,\bm g\,$ is the usual Newtonian gravity field and $\,\bm b\,$ is the gravito-magnetic field.  In terms of the metric perturbation
we have:
\begin{equation}\label{g003}
g_i=-\frac{\kappa}{2}\frac{\partial h_{00}}{\partial x^i}\qquad\textrm{and}\qquad 
b_i=-\kappa\left(\frac{\partial h_{0k}}{\partial x^j}-\frac{\partial h_{0i}}{\partial x^k}\right),
\end{equation}
where $i,\,j$ and $k$ indicate the spatial components of the space-time index. Using the anti-symmetric Levi-Civit\`a symbol 
$\epsilon_{ijk}\,$, one  may arrange the vector fields in a tensor form that is analogous to covariant electrodynamics, such as
\begin{equation}
f_{0i}=g_i\qquad\textrm{and}\qquad f_{ij}=-\frac{1}{2}\epsilon_{ijk}b_k.
\end{equation}
The analogy is limited because the components of $f_{\mu\nu}$ are in fact the $\mu=0$ component of a third rank tensor 
\cite{Campbell:1970ww,Campbell:1973grw,Campbell:1976ltw,Campbell:1977jf,Ramos:2010zza}, and thus the results 
are indeed not covariant in the same sense of the electro-dynamical formulation. 

In spite of this, the results of GEM indicate that an analogy between the formulations of gravitation and electrodynamics may exist, and 
several proposals for understanding it have emerged. In this article
we propose a modified version of Newtonian gravity where the electro-dynamical analogy emerges as a natural consequence. A similar proposal 
is presented
in \cite{Behera:2018prr}, but the tensor formulation that will be presented in this article is clearly different. We can describe our results as
an attempt to formalize the theory in a simple way, something that can benefit future studies of the subject.

The article is organized as follows: in Section \ref{I} we propose a modified Newtonian gravity field with an emerging Lorentz-like force law. 
In section \ref{P} this field is written using a tensor formulation, while in Section \ref{PP} this formulation is extended using
a 4-vector potential. Section \ref{A} repeats the preceding section results for an alternative force law, while   Section \ref{C} rounds off 
the article with our conclusions and directions for future research.

\section{\;\sc Self-interacting Newtonian gravity \label{I}}

Newtonian gravity  can be modified to obtain alternative models of gravity.  We quote 
\cite{Milgrom:1983ca,Milgrom:2014usa} as a recent proposal of this kind, but more radical possibilities have been  considered in
\cite{Fischbach:1999bc}. In our proposal the a mass density $\rho$, the matter flux density vector $\bm p$ and the gravitational vector field $\,\bm g\,$ are such that the following field equations hold:
\begin{equation}\label{g02}
\bm{\nabla\cdot g}=-\,4\pi\rho\qquad\qquad\mbox{and}\qquad\qquad\bm{\nabla\times g}=\frac{4\pi}{c}\bm{p}-\frac{1}{c}\frac{\partial\bm{g}}{\partial t}.
\end{equation}
We emphasize that (\ref{g02}) is not a new theory of gravity.
Our proposal is to demonstrate that the above equations describe a gravitational field that is equivalent to General Relativity at the gravito-electromagnetic precision level. We remember that the first order truncation of Einstein's equations gives Newtonian gravity, while a higher order truncation gives (\ref{g002}-\ref{g003}). In this article, this prediction of general relativity will obtained from to (\ref{g02}) after imposing the covariant formalism.

 The first order truncation gives Newtonian gravity, while
a higher order truncation gives (\ref{g002}-\ref{g003}), which will be shown to be equivalent to (\ref{g02}) with constraints. Let us then consider the gravitational force $\bm F$ as
\begin{equation}\label{g03}
\bm F=m\bm g+\frac{1}{c}\bm{p\times g},
\end{equation}
 where $\bm p$ is the matter flux of the source of the field.
This force law is inspired by (\ref{g002}), which is obtained from Einstein's equations, and
it is possible to flip the signal of $\,\bm{p\times g},\,$ a possibility that is entertained in Section \ref{A}. The aim of the present 
article is to examine the consistency of the above proposal and interpret it physically. 

In which context would be relevant the proposal (\ref{g02}-\ref{g03})? Probably not in the context ascertained by the 
gravity probe B experiment (GP-B) \cite{Everitt:2015qri}, where kinematic effects are irrelevant. In more general contexts, the
source of the field can be endowed with high linear momentum, high spin or another time dependence, and we hypothesize that the proposal
can be suitable in these cases. The model may be considered as self-interacting because the
time variation of gravity also contributes to gravity itself, however in Section \ref{PP} we show that linear Maxwell-like field 
equations may be obtained from (\ref{g02}). Let us start the characterization of the model questioning why
the $\,\bm{p\times g}\,$ contribution to $\bm F\,$ is difficult to observe. Using the international system of units, where $G$ is the 
universal coupling constant of gravity, we form
\begin{equation}
\bm g \to\frac{4\pi}{G}\bm g,\qquad\textrm{and we define}\qquad \frac{G}{H}=c,
\end{equation}
where we call $\,H\,$ a dynamic coupling constant. In these units the field equations are
\begin{equation}\label{g04}
\bm{\nabla\cdot g}=-\,G\rho\qquad\qquad\mbox{and}\qquad\qquad\bm{\nabla\times g}=H\bm{p}-\frac{H}{G}\frac{\partial\bm{g}}{\partial t},
\end{equation}
and the gravitational force is
\begin{equation}
\bm F=\frac{1}{4\pi}\Big(G m\bm g\,+\,H\bm{p\times g}\Big).
\end{equation}
Finally, we evaluate the intensity of the dynamical coupling, so that
\begin{equation}\label{g05}
\frac{G}{H}=c,\qquad G=6.674\times 10^{-11} \frac{m^3}{kg\cdot s^2}\qquad\Rightarrow \qquad 
H=2.226\times 10^{-19}\frac{ m^2}{kg\cdot s}.
\end{equation}
Consequently the dynamical interaction is extremely weak compared to that of static gravity. If this coupling is in fact physical, 
its weakness explains why it is hardly observable. This situation is different from electrodynamics, where the coupling of the magnetic force 
is also weak compared to the coupling of the electrostatic interaction. However, these interactions are observed separately, something that it is technically more difficult to do here. 

 Let us now go back to the cgs system of units that will be used throughout the article. The continuity equation is obtained after calculating the divergence of the curl of the 
gravitational vector field (\ref{g02})
\begin{equation}\label{g06}
\frac{\partial\rho}{\partial t}+\bm{\nabla\cdot p}=0,
\end{equation}
which indicates the conservation of mass, in the same way that the electric charge is conserved in electrodynamics. 
A reasonable interpretation of (\ref{g06}) is that as the electromagnetic interaction does not destroy the electric charge,
the gravitational interaction also does not destroy mass.
Another simple consequence of (\ref{g02}) comes from the scalar product between the curl of gravity and the gravity vector, so that
\begin{equation}\label{g07}
\frac{1}{8\pi c}\frac{\partial |\bm g|^2}{\partial t}+\frac{1}{4\pi}\bm{g\cdot\nabla\times g}=\frac{1}{c}\bm{g\cdot p}.
\end{equation}
We observe that $\, \bm{g\cdot\nabla\times g}=0\,$ if the field is conservative, and we are not certain that this is so, therefore
we keep this term in the equation and eventually impose the nullity of the second term as a constraint to the theory.
The electro-dynamical analogue of (\ref{g07}) is the Poynting theorem. In this case, it is not possible to get a gravitational Poynting 
vector because $\bm{g\times g}=0$. Despite this, we can understand $|g|^2/8\pi$ as the energy density of the gravitational field, and 
$\bm{g\cdot p}$ is the density of the work done by the field over the mass of the system. If the mass 
moves away from the source, $\bm{g\cdot p}<0$ and the total energy density diminishes. It is possible to formulate GEM in such a way that 
the signal of $\,\bm{p\cdot g}\,$ in (\ref{g07}) is flipped \cite{Vieira:2016csi}, but we consider this proposal unphysical. 
The interpretation of $\bm{g\cdot\nabla\times g}$  is  similar to that of $\bm{g\cdot p}$, but the energy change is instead associated to the 
self-interaction of the field. We are unable to obtain an expression for the conservation of the linear
momentum in the same way as is done for electro-dynamics, and thus the field equations, force law, conservation of mass and 
the conservation of energy are the only results that we obtain from the model in this formulation. In the next section, we try a tensor 
formulation in order to shed further light on the physical model.

\section{\;\sc Tensor formulation using the field approach \label{P}}
In order to obtain a covariant theory, we take the electrodynamic field tensor as a model to propose the gravitational field tensor
\begin{equation}\label{p03}
C_{\mu\nu}\,=\,\left[
\begin{array}{llll}
0   & -g_1       & -g_2       & -g_3\\
g_1 & \;\;\;0    & \;\;\,g_3  & -g_2\\
g_2 & -g_3       & \;\;\,0    & \;\;\,g_1 \\
g_3 & \;\;\,g_2  & -g_1       & \;\;\,0
\end{array}
\right]\qquad\mbox{where}\qquad\bm g=\big(g_1,\,g_2,\,g_3\big),
\end{equation}
 Accordingly,
\begin{equation}
C_{i0}=g_i,\qquad\qquad C_{ij}=\,\epsilon_{ijk} g_k.
\end{equation}
In order to describe this dynamical gravity in terms of an energy-momentum tensor, we introduce the symmetric tensor $\,\tau^{\mu\nu}\,$ 
\begin{equation}\label{p14}
\tau_{\mu\nu}=\tau^{\mu\nu}=\left\{
\begin{array}{l}
1\qquad\mbox{for}\qquad \mu=\nu,\\
0\qquad\mbox{for}\qquad\mu\neq\nu.
\end{array}
\right.
\end{equation}
Although it is not a tetrad, this tensor can be understood as the square root of the
Minkowskian metric tensor because $\,\eta^{\mu\nu}=\tau^{\mu\kappa}\tau_\kappa^{\;\;\nu}.\,$ 
Thus, the field tensor allows us to rephrase the gravity field equations (\ref{g02}) as
\begin{equation}\label{p04}
\tau^\mu_{\;\;\mu'}\tau^\nu_{\;\;\nu'}\partial_\nu C^{\mu'\nu'}=\frac{4\pi}{c}p^\mu,\qquad\mbox{where}\qquad p^\mu=\big(c\rho,\,\,\bm p\big).
\end{equation}
$\,x^\mu=\big(ct,\,\bm x\big)\,$ is the contravariant coordinate $4-$vector and $\,p^\mu\,$ and is the contravariant $4-$vector momentum 
density, that we could also call the $4-$vector matter current. The continuity equation (\ref{g06}) is simply
\begin{equation}\label{p05}
\partial_\mu p^\mu=0,
\end{equation}
and the  covariant expression  for the gravitational force is also obtained, where
\begin{equation}\label{p08}
\frac{d p^\mu}{dt}=\frac{1}{c}C^{\nu\mu}p_\nu.
\end{equation}
The spatial components of (\ref{p08}) give the gravitational force and the $\mu=0$ component gives
\begin{equation}\label{p09}
c^2\frac{d\rho}{dt}=\bm{g\cdot p},
\end{equation}
which describes the variation of the total energy density of the system. The anti-symmetry of the field tensor imposes
\begin{equation}\label{p10}
p^\mu\frac{d p_\mu}{dt}=\frac{1}{c}C_{\mu\nu}p^\mu p^\nu=0\qquad\mbox{thus}\qquad \frac{d}{dt} \big(p^\mu p_\mu\big)=0,
\end{equation}
and therefore $p_\mu p^\mu$ is a constant. It is natural to associate this constant to the rest energy density $E$, and consequently we interpret the four-momentum vector (\ref{p04}) relativistically, so that
\begin{equation}\label{p11}
p_\mu p^\mu\,=\,\rho^2c^2-\bm{p\cdot p}=\frac{E^2}{c^2}
\end{equation}
The above result permits interesting physical interpretations.  Now, we rephrase the equations of motion (\ref{p04}) as
\begin{equation}\label{p15}
\partial_\mu C_{\nu\kappa}+\partial_\nu C_{\kappa\mu}+\partial_\kappa C_{\mu\nu}=
\frac{4\pi}{c}\epsilon_{\mu\nu\kappa\lambda} p^\lambda
\end{equation}
where $\,\epsilon_{\mu\nu\kappa\lambda}\,$ is the anti-symmetric Levi-Civit\`a symbol. One may argue that the introduction of 
$\tau_{\mu\nu}$ could be suppressed by redefining several vectors, for example 
$\,\partial_\mu\to\partial_\lambda\tau^\lambda_{\;\;\mu}=(\partial_0,\,-\bm\nabla).\,$ We understand that these definitions are unnatural 
and complicate the comparison with the alternative formulation of Section $\ref{A}$, which would be easier if the $4-$vectors are common to
both of the formulations. Combining (\ref{p04}) and (\ref{p08}), we obtain
\begin{equation}\label{p16}
\frac{d p_\mu}{dt}=\frac{1}{4\pi}\tau^{\nu\nu'}\tau^{\lambda\lambda'}\,C_{\mu\nu}\partial_\lambda C_{\lambda'\nu'},
\end{equation}
and additionally combining (\ref{p15}-\ref{p16}) produces
\begin{equation}\label{p17}
\frac{dp^\mu}{dt}=\partial_\kappa T^{\kappa\mu}+S^\mu.
\end{equation}
Here $T_{\mu\nu}$ is the energy-momentum tensor and  $\,S_\mu\,$ is the source term, explicitly
\begin{equation}\label{p18}
T^{\mu\nu}=\frac{1}{4\pi}\left(\tau_{\;\;\kappa}^\nu C^{\mu\sigma}C^{\kappa\sigma'}-\frac{1}{4}\eta^{\mu\nu}\tau_{\kappa\kappa'} C^{\kappa'\sigma'}C^{\kappa\sigma}\right)
\tau_{\sigma\sigma'},\qquad
S^\mu=\frac{1}{2c}\eta^{\mu\mu'}\epsilon_{\mu'\nu\lambda\kappa}\,p^\kappa \tau^{\nu\nu'}\tau^{\lambda\lambda'}C_{\nu'\lambda'}.
\end{equation}
However, explicit calculations shows that
\begin{equation}\label{p19}
T^{\mu\nu}=0\qquad\qquad
S^\mu=\left(\frac{\bm{g\cdot p}}{c},\,\frac{d\bm p}{dt}\right).
\end{equation}
and thus (\ref{p17}) is a trivial identity. This identically zero energy-momentum tensor presents an interesting difference to electromagnetism. Despite its consistency, the field formulation seems not to be useful for further developments, but we will see that it is useful for the physical interpretation.  In the next section we try a potential formulation in order to obtain a more interesting results in terms of conservation laws and  a relativistically description.

\section{\; \sc The tensor formulation using the potential approach \label{PP}}
The gravitational field can be  represented by the scalar potential $\Phi$ and by the vector potential $\bm\Psi$, so that
\begin{equation}\label{p001}
\bm g=-\bm\nabla\Phi-\frac{1}{c}\frac{\partial\bm\Psi}{\partial t}-\bm{\nabla\times\Psi}.
\end{equation}
Consequently the field equations (\ref{g02}) become
\begin{align}
\nonumber
\nabla^2\Phi\,+\,\frac{1}{c}\frac{\partial}{\partial t}\big(\bm{\nabla\cdot\Psi\big)}&\;=\;4\pi\rho\\
\label{p002}
\nabla^2\bm\Psi-\bm\nabla\big(\bm{\nabla\cdot\Psi}\big)&\;=\;\frac{4\pi}{c}\bm p\,+\,\frac{1}{c}\frac{\partial}{\partial t}\bm\nabla\Phi\,+\,\frac{1}{c^2}\frac{\partial^2\bm\Psi}{\partial t^2}+\frac{2}{c}\frac{\partial}{\partial t}\bm{\nabla\times\Psi}.
\end{align}
In this formulation, what we called the gravity field $\bm g$ is in fact composed by two more fundamental fields, the  gravito-electric field $\bm g_E$ and the gravito-magnetic field $\bm g_B$. This hypothesis replaces the electric and magnetic contributions to gravity discussed in section \ref{I} and permits the interpretation of gravity as composed of a field generated by static mater $\bm g_E$ and another generated by moving matter $\bm g_B$.
Nevertheless, a simpler description is obtained by defining
\begin{equation}\label{p003}
\bm g=\bm g_E-\bm g_B,\qquad\qquad\textrm{where}\qquad\qquad\bm g_E=-\bm\nabla\Phi-\frac{1}{c}\frac{\partial\bm\Psi}{\partial t}\qquad\mbox{and}\qquad\bm g_B=\bm{\nabla\times\Psi}.
\end{equation}
Therefore, using (\ref{p003}) in (\ref{g02}) we obtain
\begin{align}
\nonumber&\bm{\nabla\cdot g}_E=-\,4\pi\rho\;\;\;\,\qquad\qquad\qquad \bm{\nabla\cdot g}_B=0\\
\label{p004}&\bm{\nabla\times g}_E=\frac{1}{c}\frac{\partial\bm g_B}{\partial t}\qquad \qquad\qquad\,
\bm{\nabla\times g}_B=-\frac{4\pi}{c}\bm p+\frac{1}{c}\frac{\partial\bm g_E}{\partial t},
\end{align}
a set of equations very similar to the Maxwell electromagnetic field equations.  We stress that these field equations are different from previous approaches to GEM 
\cite{Campbell:1970ww,Campbell:1973grw,Campbell:1976ltw,Campbell:1977jf,Ramos:2010zza}. The important point is  that the signals of the curl and of the time derivative agree in (\ref{p004}), 
while there is a single flipped signal in the previous articles, thus it is not possible
to couple the fields like $\bm g=\bm g_E-\bm g_B$ to obtain (\ref{g02}). The entire novelty of the results presented in this article depends on these field equations. The conservation of the energy in terms of the energy
density $\,u\,$ and the momentum flux $\,\bm s\,$ is
\begin{equation}\label{ap05}
\bm{\nabla\cdot s}+\frac{\partial u}{\partial t}=\bm{p\cdot g}_E,\qquad\mbox{where}\qquad \bm{s}=c\frac{\bm g_E\bm{\times g}_B}{4\pi}
\qquad u=\frac{|\bm g_E|^2-|\bm g_B|^2}{8\pi}.
\end{equation}
The only differences are two signal flips, which is how the field got its name: gravito-electromagnetism. At this point the gravitational formulation deviates from the electromagnetic formalism, and the gravitational potential second rank tensor is
\begin{equation}\label{p005}
\mathscr{C}^{\mu\nu}\,=\,-\,\Big(\partial^\mu Q^\nu-\partial^\nu Q^\mu\Big)\qquad
\textrm{where}\qquad Q^\mu=\big(\,\Phi,\,\bm\Psi\,\big)
\end{equation}
is the gravitational potential $4-$vector. From this, we immediately obtain that
\begin{equation}\label{p006}
\mathscr C_{i0}=\big(g_E\big)_i\qquad\mbox{and}\qquad \mathscr C_{ij}=\epsilon_{ijk}\big(g_B\big)_k
\end{equation}
As already mentioned in the previous section, we could redefine $\partial_\mu$ and $Q_\mu$ and absorb the $\tau^{\mu\nu}$ tensor. 
This has not been done because we want identical $4-$vectors for the alternative formulation of the theory in Section \ref{A}, which will
make the similarities between both of the theories much clearer.
The potential field tensor (\ref{p005}-\ref{p006}) permits us to recover the equations of motion (\ref{p002}) from
\begin{equation}\label{p007}
\partial_\nu\mathscr{C}^{\nu\mu}=\frac{4\pi}{c} p^\mu.
\end{equation}
In the same manner as in electrodynamics, equation (\ref{p007}) only contains the non-homogeneous relations of (\ref{p004}), and the homogeneous equations are obtained from the identity
\begin{equation}\label{p008}
\partial_\lambda\Big( \tau^\lambda_{\;\;\mu}\mathscr{C}_{\nu\kappa}+\tau^\lambda_{\;\;\nu}\mathscr{C}_{\kappa\mu}+\tau^\lambda_{\;\;\kappa}\mathscr{C}_{\mu\nu}\Big)=0.
\end{equation}
For the $4-$vector momentum density, we write
\begin{equation}\label{p009}
\frac{d p^\mu}{dt}=\frac{1}{c}\,\tau^\mu_{\;\;\mu'}\tau^\nu_{\;\;\nu'}\mathscr{C}^{\mu'\nu'}p_\nu,
\end{equation}
meaning that
\begin{equation}\label{p010}
\frac{dp^0}{dt}=\frac{1}{c}\bm{p\cdot g}_ E\qquad\qquad\textrm{and}\qquad\qquad\frac{d\bm p}{dt}=\rho\,\bm g_E-\frac{1}{c}\bm{p\times g}_B.
\end{equation}
This formulation of gravito-electromagnetism appears to be the same as the Heaviside gravity from \cite{Behera:2018prr}, but some signal
flips have to be understood  in order to comprehend the exact relationship between the formulations. On the other hand,
comparing (\ref{p010}) to (\ref{g03}) and (\ref{p09}), two constraints are obtained, leading to
\begin{equation}\label{p011}
 \bm{p\cdot g}_B=0;\qquad c\rho\bm g_B-\bm{p\times g}_E=\bm 0,\qquad\textrm{or equivalently}
 \qquad\bm g_E\bm{\cdot g}_B=0.
\end{equation}
The gravitational force vector $\,d\bm p/dt\,$ is thus coplanar to $\bm p\,$ and $\,\bm g_E.\,$ The existence of the constraint (\ref{p011})
 indicates that our model is not equivalent to the Heaviside gravity formulation of \cite{Behera:2018prr}, and it again different from
electromagnetism. We notice that the force law in (\ref{p011}) fits exactly with (\ref{g002}), and the flipped signal
may be obtained from redefining $\,\bm b\,$ in (\ref{g003}). However, as we have already pointed out, it is not possible to consider an exact
match between both of these expressions because (\ref{g002}) comprises the $\,\mu=0\,$ component of a presently unknown third rank tensor.
From the force law, we obtain the momentum conservation inside a region of volume $V$ of surface $S$, 
\begin{equation}\label{p012}
\frac{d \bm p_P}{dt}+\frac{d\bm p_F}{dt}=\sum_{j=1}^3\oint_S \mathscr T_{ij}n_jda
\end{equation}
where $\,\bm p_P\,$ is the momentum contained in the particles and the unit vector $\,\bm n\,$ is normal to the integration region. The momentum $\,\bm p_F\,$ contained in the fields and the stress tensor $\,\mathscr T_{ij}\,$
are
\begin{equation}\label{p013}
\bm p_F=\frac{1}{c^2}\int_V \bm s\, dv\qquad\mbox{and}\qquad \mathscr T_{ij}=\frac{1}{4\pi}\Big[(g_B)_i(g_B)_j-(g_E)_i(g_E)_j\Big]+ u\,\delta_{ij}
\end{equation}
We are ready to obtain the expression of the force law in terms of a covariant divergence. Using (\ref{p008}), we obtain the identity
\begin{equation}\label{p014}
\tau_\mu^{\;\;\mu'}\tau_\nu^{\;\;\nu'}\mathscr C^{\kappa\nu}\,\partial_\kappa\mathscr C_{\mu'\nu'}=
\frac{1}{4}\partial_\mu\Big(\tau_\kappa^{\;\;\kappa'}\tau^\nu_{\;\;\nu'}\mathscr C^{\kappa\nu}\mathscr C_{\kappa'\nu'}\Big).
\end{equation}
Using (\ref{p007}), (\ref{p009}) and (\ref{p014}), we obtain the force law as a covariant divergence
\begin{equation}\label{p015}
\frac{dp^\mu}{dt}=\partial_\kappa\mathscr{T}^{\kappa\mu},
\end{equation}
where the energy-momentum tensor is
\begin{equation}\label{p016}
\mathscr{T}^{\mu\nu}=\frac{1}{4\pi}\left(\tau^{\mu\mu'}\tau^{\kappa'}_{\;\;\kappa}\mathscr{C}_{\mu'\kappa'}\,\mathscr{C}^{\nu\kappa}\,-\,\frac{1}{4}\eta^{\mu\nu}\,\tau^{\kappa'}_{\;\;\kappa}\,\tau^{\lambda'}_{\;\;\lambda} \mathscr{C}_{\kappa'\lambda'}\mathscr{C}^{\kappa\lambda}\right).
\end{equation}
Explicitly, the components of the energy-momentum tensor are
\begin{align}\label{p017}
 &\mathscr T_{00}=\frac{1}{8\pi}\left(\big|\bm g_E\big|^2-\big|\bm g_B\big|^2\right)\qquad\qquad\quad\;\,
 \mathscr T^{ii}\,=\,\mathscr T^{00}+\frac{1}{4\pi}\left[\;\big(\bm g_B\big)_i^2-\big(\bm g_E\big)_i^2\;\right]\\
 \nonumber
 &\mathscr T^{0i}=\frac{1}{4\pi}\Big(\bm g_E\times\bm g_B\Big)_i=\,-\,\mathscr T^{i0}
 \qquad\qquad\mathscr T^{ij}=\frac{1}{4\pi}\left[\,\big(\bm g_B\big)_i\big(\bm g_B\big)_j-\big(\bm g_E\big)_i\big(\bm g_E\big)_j\,\right].
\end{align}
Accordingly,
\begin{equation}\label{p018}
\nonumber\mathscr T_{\mu\nu}\tau^{\mu\nu}=0,\qquad
\mathscr T_\mu^{\;\;\mu}=2\,\mathscr T_{00},\qquad
\mathscr T_{\mu\nu}\mathscr T_{\mu\nu}\tau^{\mu'\mu'}\tau^{\nu\nu'}=2\Big(\mathscr T_{00}^2+|\mathscr T_{0i}|^2\Big)\qquad\mbox{and}\qquad
\mathscr T_{\mu\nu}\mathscr T^{\mu\nu}=2\Big(\mathscr T_{00}^2-|\mathscr T_{0i}|^2\Big).
\end{equation}
The nullity of $\,\mathscr T_{\mu\nu}\tau^{\mu\nu}\,$  indicates that an invariant property plays a role here that is played identically by the nullity  of the trace of the electromagnetic 
energy-momentum tensor. This particular feature justifies the introduction of $\,\tau^{\mu\nu}\,$ to the formulation of gravito-electromagnetic theory presented in this article, but the 
physical meaning of $\tau^{\mu\nu}$ is a matter for future research. Finally, from (\ref{p012}) we obtain
\begin{equation}\label{p0170}
 \frac{dp^0}{dt}=\frac{\partial u}{\partial t} +
  \bm{\nabla\cdot s}.
\end{equation}
This result connects is a consistency test that links the field formulation (\ref{p17}) to the potential formulation (\ref{ap05}), as expected. Conversely, the spatial components of (\ref{p012}) generate the Lorentz force without adding new constraints, 
and thus the consistency of the formulation is demonstrated. The Lagrangian density of this formulation of GEM is simply
\begin{equation}\label{p0180}
\mathcal{L}=-\frac{1}{16\pi}\mathscr C_{\mu\nu}\mathscr C^{\mu\nu}\,+\frac{1}{c}p_\mu Q^\mu,
\end{equation}
and (\ref{p007}) is easily obtained from (\ref{p018}).

\section{\sc The alternative gravity law \label{A}}
In this section, we study the gravito-electromagnetic force
\begin{equation}\label{a01}
\bm F=m\bm g-\frac{1}{c}\bm{p\times g},
\end{equation}
which will be shown to be coherent to the field equations
\begin{equation}\label{a02}
\bm{\nabla\cdot g}=-\,4\pi\,\rho,\qquad\bm{\nabla\times g}=-\frac{4\pi}{c}\bm p+\frac{1}{c}\frac{\partial\bm g}{\partial t}.
\end{equation}
In this formulation, the field tensor components are
\begin{equation}\label{a03}
C_{i0}=g_i;\qquad C_{ij}=-\epsilon_{ijk}g_k,
\end{equation}
and equations (\ref{p04}) and (\ref{p08}) hold. On the other hand, (\ref{p15}) flips a signal
\begin{equation}\label{a04}
\partial_\mu C_{\nu\kappa}+\partial_\nu C_{\kappa\mu}+\partial_\kappa C_{\mu\nu}=
-\frac{4\pi}{c}\epsilon_{\mu\nu\kappa\lambda} p^\lambda.
\end{equation}
The equation of motion in terms of a energy-momentum tensor is identical to (\ref{p17}), where the components of the energy-momentum tensor $\,T_{\mu\nu}\,$ are identically zero, but the source term flips its signal because of (\ref{a04}). 
Hence, both of the formulations are equally consistent, and the correct physical expression must be obtained through  experimental tests 
of  (\ref{g03}) and (\ref{a01}). The potential formulation of the alternative law is obtained for
\begin{equation}\label{a06}
\bm g=\bm g_E+\bm g_B,\qquad\qquad\textrm{where}\qquad\qquad\bm g_E=-\bm\nabla\Phi+\frac{1}{c}\frac{\partial\bm\Psi}{\partial t}\qquad\mbox{and}\qquad\bm g_B=\bm{\nabla\times\Psi}.
\end{equation}
and the gravito-electrodynamical field equations are
\begin{align}
\nonumber&\bm{\nabla\cdot g}_E=-\,4\pi\rho\;\;\;\,\qquad\qquad\qquad \bm{\nabla\cdot g}_B=0\\
\label{a07}&\bm{\nabla\times g}_E=\frac{1}{c}\frac{\partial\bm g_B}{\partial t}\qquad \qquad\qquad
\bm{\nabla\times g}_B=-\frac{4\pi}{c}\bm p+\frac{1}{c}\frac{\partial\bm g_E}{\partial t}.
\end{align}
Additionally,
\begin{equation}\label{a08}
\mathscr C_{\mu\nu}=-\,\tau_\mu^{\;\;\mu'}\tau_\nu^{\;\;\nu'}\Big(\partial_{\mu'}Q_{\nu'}-\partial_{\nu'}Q_{\mu'}\Big),
\end{equation}
leads to,
\begin{equation}\label{a09}
\mathscr C_{i0}=\big(g_E\big)_i,\qquad\qquad \mathscr C_{ij}=\epsilon_{ijk}\big(g_B\big)_k,
\end{equation}
and equations (\ref{p007}-\ref{p008}) are immediately recovered. From (\ref{p009}), we produce
\begin{equation}\label{a10}
\frac{dp^0}{dt}=\frac{1}{c}\bm{p\cdot g}_ E\qquad\qquad\textrm{and}\qquad\qquad\frac{d\bm p}{dt}=\rho\,\bm g_E+\frac{1}{c}\bm{p\times g}_B.
\end{equation}
Accordingly, two constraints are obtained, so that
\begin{equation}\label{a11}
 \bm{p\cdot g}_B=0;\qquad c\rho\bm g_B+\bm{p\times g}_E=\bm 0,\qquad\textrm{or equivalently}
 \qquad\bm g_E\bm{\cdot g}_B=0.
\end{equation}
In summary, the results of the previous formulation and the present alternative formulation are related by a symmetry that
may be expressed in several ways, such as
\begin{equation}\label{a12}
\bm g_B\to-\bm g_B,\qquad\qquad\textrm{or}\qquad\qquad \bm\Psi\to-\bm\Psi\qquad\qquad\textrm{or}\qquad\qquad Q^\mu\to Q^\nu\tau_\nu^{\;\;\mu}. 
\end{equation}
Thus, under the alternative gravity law the equivalents of (\ref{p012}-\ref{p017}) are immediately obtained using (\ref{a12}). 
The most important difference is the signal flip in the ``Pointing vector'' of (\ref{p017}), which means that
 the momentum flux is reversed in both of the formulations. In \cite{Behera:2018prr}, it is concluded that both of the approaches are 
equivalent. We do not exclude this hypothesis, but we believe that  further research is necessary in order to verify its plausibility. 
On the other hand, we are not certain that both  approaches are identical because there are several flipped signals between the 
field equations presented here and those presented in  table $2$ of \cite{Behera:2018prr}.

\section{\sc Conclusion\label{C}}

In this paper, a proposal of the gravity field equations (\ref{g03}) was covariantly organized in a potential approach 
(\ref{p003}-\ref{p004}) that is alternative to the usual formulation of the gravito-electromagnetism. However, covariant approaches to GEM were already obtained in several cases \cite{Behera:2018prr}, and one may question the necessity of a further formulation. The originality of the present theory concerns the linear decomposition of the gravity field in terms of the static matter field $\bm g_E$ and the moving matter field $\bm g_B$, something that is not possible in the previous formulations. Hence the article presents  two novelties, a conceptual idea that unified the gravito-electic and gravito-magnetic fields into a single gravity field, and the technical way to construct such covariant theory, that is different from the usual GEM formulations.
This new covariant formulation is in fact more difficult to be obtained, and the main technical point of the analysis is the introduction of the $\tau^{\mu\nu}$ tensor (\ref{p14}), and this demonstrates the originality of the results presented in this article when compared to the 
electromagnetic formulation.

A further motivation for this article concerns the fact  that formal expressions are fundamental when investigating the solutions of a theory, and also for building models that can be tested 
experimentally. Therefore, researching classical solutions is a clear and important direction for future research. An immediate and interesting
direction for future research concerns the gravitational waves from this field theory, and their relation to the usual gravitational waves predicted
by general relativity. Another important direction for future research concerns whether these results are useful for describing  experimental data different to those presently coming from the gravity probe B experiment (GP-B) \cite{Everitt:2015qri}, or proposing a different collection of data, perhaps with higher precision, perhaps in situations where the gravity source may experience momentum, spin or 
time-dependence.  However, many former results of GEM can be studied using the presented formalism, and we quote the previous works on 
gravitational waves \cite{Campbell:1976ltw}, gauge structure \cite{Campbell:1973grw} and  Lagrangian formulation\cite{Ramos:2010zza} as interesting references for these future directions of research.

Another possibility 
concerns the quantization of gravito-electromagnetism, which is almost identical to the usual quantization of the electromagnetic field with
 the most important difference being the
 constraint $\,\bm g_E\bm{\cdot g}_B=0.\,$ However, the most serious problem of this approach is conceptual: the existence
of an attractive gravity force whose quantum particle is a graviton of the spin$-1$. We hope that the formalism presented here 
will also be useful should this possibility be conceptually acceptable.

%
%
%
%

\bibliographystyle{unsrt} 
\bibliography{bib_gravidade}
\end{document}